\documentclass[twocolumn,showpacs,preprintnumbers,amsmath,amssymb]{revtex4-1}

\usepackage{graphicx}% Include figure files
\usepackage{dcolumn}% Align table columns on decimal point
\usepackage{bm}% bold math
\usepackage{float}
\usepackage{float}
\usepackage{color}
\usepackage{array}
\usepackage{rotating}
\usepackage{ulem}
\graphicspath{{FiguresMain/}}

\begin{document}

\preprint{APS/123-QED}

\title{Theoretical Analysis of an Ideal Noiseless Linear Amplifier for Einstein-Podolsky-Rosen Entanglement Distillation}
\author{J. Bernu$^{1}$, S. Armstrong$^1$, T. Symul$^{1}$, T.C. Ralph$^{2}$ P. K. Lam$^{1,3}$}

\affiliation{$^1$Centre of Excellence for Quantum Computation and Communication Technology, Department of Quantum Science, The Australian National University, Canberra, Australia.}
\affiliation{$^2$Centre of Excellence for Quantum Computation and Communication Technology, School of Mathematics and Physics, University of Queensland, St. Lucia QLD 4072, Australia.}
\affiliation{$^3$College of Precision Instrument and Opto-electronics Engineering, Key Laboratory of Optoelectronics Information Technology of Ministry of Education, Tianjin University, Tianjin, 300072, China}

\date{\today}% It is always \today, today,

\begin{abstract} 
We study the operational regime of a noiseless linear amplifier based on quantum scissors that can nondeterministically amplify the one photon component of a quantum state with weak excitation. It has been shown that an arbitrarily large quantum state can be amplified by first splitting it into weak excitation states using a network of beamsplitters. The output states of the network can then be coherently recombined. In this paper, we analyse the performance of such a device for distilling entanglement after transmission through a lossy quantum channel, and look at two measures to determine the efficacy of the noiseless linear amplifier. The measures used are the amount of entanglement achievable and the final purity of the output amplified entangled state. We study the performances of both a single and a two-element noiseless linear amplifier for amplifying weakly excited states. Practically, we show that it may be advantageous to work with a limited number of stages.
\end{abstract}

\newcommand{\red}[1]{ { \color{red}#1} }
\newcommand{\blue}[1]{ { \color{blue}#1} }
\newcommand{\ket}[1]{\left| #1 \right\rangle}
\newcommand{\bra}[1]{\left\langle #1 \right|}
\newcommand{\braket}[2]{\left\langle#1\middle| #2 \right\rangle}
\newcommand{\braopket}[3]{\left\langle#1\middle| #2 \middle| #3 \right\rangle}
\newcommand{\e}{\epsilon}
\renewcommand{\l}{\lambda}
\newcommand{\h}{\eta}
\renewcommand{\r}{\rho}
\renewcommand{\k}{\kappa}
\newcommand{\sech}{\text{sech}}
\newcommand{\EPR}{\text{EPR}}
\newcommand{\s}{ \sigma}
\renewcommand{\dag}{^\dagger}
\newcommand{\Tr}{\text{Tr}\;\!}
\newcommand{\NLA}{\text{NLA}}

\maketitle

%\section{Introduction}

Light possesses unique quantum properties that enable the realisation of quantum information protocols that have no classical equivalent. Amongst these protocols are the realisation of quantum computers \cite{Qcomp} that can solve various problems much more efficiently than their classical equivalents, or the implementation of Quantum Key Distribution devices \cite{Gis02, Gro03, Lan05} that enable absolutely secure communication ensured by physical laws \cite{Sca09}. For these new devices to achieve their full potential, a quantum internet will need to be developed \cite{Kim08}, which is a network able to support the efficient communication of quantum states. Most, if not all, quantum communication protocols rely on distributed entanglement \cite{EPR}. Entanglement may be treated as a physical resource, somewhat like energy, associated with the peculiar non-classical correlations that are possible between physically separated quantum systems. For example, entanglement distributed between two parties allows for quantum teleportation \cite{Vai94,Fur98,Bow03} whereby a quantum state can be transferred over a distance via a classical communication channel.

Quantum communication, however, rapidly decoheres over real-world distances due to losses and noise on the quantum channel. Therefore a device such as a quantum repeater is required to extend the communication range to more practical distances. Construction of such a device represents a major challenge. While some experimental implementations have been proposed \cite{San11}, presently the best known method for constructing a quantum repeater is to concentrate or distil a small amount of useful entanglement from a large amount of decohered entanglement. The distilled entanglement can then be used as a resource for teleporting quantum information.

A promising branch of quantum communication research is in the so called continuous-variable (CV) regime. In the CV regime, the degrees of freedom of the quantum system used to encode the quantum information have a continuous eigenvalue spectrum, i.e. measurement outcomes are not quantized, as opposed to discrete-variable (DV) systems where only discrete eigenvalues are allowed. In quantum optics, CV systems are usually measured using highly efficient homodyne detection schemes that give a real value outcome, while DV systems rely on photon counters that give an integer value outcome. One advantage of CV quantum communication over DV is the potential ability to achieve a high effective bandwidth by encoding the quantum information at large side-band frequencies around the optical carrier. 

It is not possible to distil a Gaussian state, which is the workhorse resource state for CV systems, using only Gaussian operations \cite{Gie02,Eis02,Fiu02}. This poses a significant hurdle in building a quantum repeater for CV systems. Only in specific circumstances where the decoherence process itself produces non-Gaussian states is distillation possible using Gaussian operations. Such a scenario has been investigated by Franzen {\it et al.} on squeezed states \cite{Fra06}. For a generic distillation protocol, one needs to 'break' the Gaussian character of CV systems by, for example, introducing DV elements such as photon counting.

Two different schemes have been proposed to realise such a hybrid (CV and DV) quantum repeater. Both schemes rely on the implementation of non-deterministic Noiseless Linear Amplifiers (NLA) to perform entanglement purification \cite{Ral09}. One scheme, proposed by Fiurasek \cite{Fiu09}, relies on multiple photon addition and photon subtraction operations to engineer arbitrary quantum operations, especially NLA-based operations. The scheme was subsequently demonstrated using a simplified version that approximated an NLA with a gain of 2 \cite{Zav11}. 

Another approach was proposed by Ralph and Lund based on the parallelisation of quantum scissors \cite{Ral09}, and was experimentally demonstrated for a single stage \cite{Xia10,Fer10}. In this approach each quantum scissor will only produce a linear superposition of zero and one-photon states. Therefore a coherent linear network of beam-splitters is necessary in order to decompose large quantum states into many smaller states that will be individually amplified. An extension of the quantum scissors scheme has been proposed where each single stage can produce states containing up to two photons \cite{Jef10}, slightly reducing the reliance of this approach on the decomposition of large states.

The photon addition and subtraction method, on the other hand, can act on any arbitrary state, at the expense of a more complex setup. We note that new post-selection schemes to circumvent the experimental complexity of physically implementing noiseless linear amplification have been proposed \cite{PostSelec1, PostSelec2} and recently demonstrated \cite{PostSelec3}, but there are restrictions to their applications.

In this paper we focus on the probabilistic NLA proposed by Ralph et Lund \cite{Ral09}. In Section I we briefly review the basic ideas behind entanglement distillation using the NLA. In Section II we define benchmarks to determine when the process is experimentally useful. In Section III we study the ability of a single stage NLA to surpass these benchmarks. In Section IV we study the dual stage case and compare it to the single stage case. In section V we investigate using a high number of stages. Finally, we conclude in Section VI.

\section{The Noiseless Linear Amplifier}

\begin{figure}[!!h]
%\begin{tabular}{cc}
%{\bf a)}\\
%&\includegraphics[width=7cm]{NLASetup1.pdf}\\
%\\
%{\bf b)}\\
%&\includegraphics[width=7cm]{NLASetup2.pdf}\\
%{\bf c)}\vspace{.2cm}\\
%&\includegraphics[width=7cm]{DistNLA.pdf}
%\end{tabular}
\includegraphics[width=7cm]{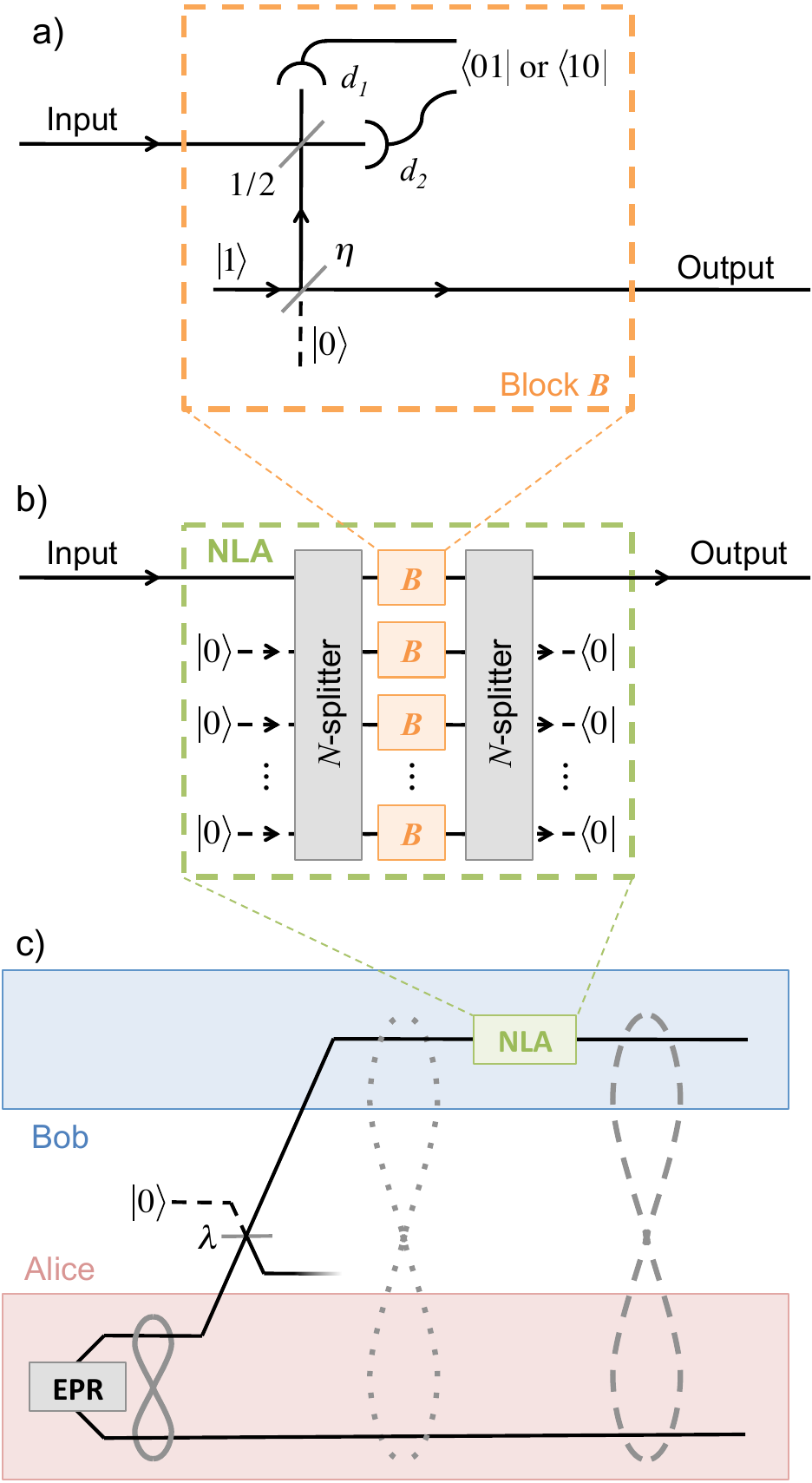}
\caption[NLA]{ {\bf a)} Schematics of the building blocks of the general NLA. $d_1$ and $d_2$ are single photon detectors, $1/2$ and $\eta$ represent beamsplitters with respectively 50\% and $\eta$ transitivity. {\bf b)} Schematics of the multi-block NLA. The eventually bright input beam is split into $N\ll1$ weakly excited beams each going into one amplification block. When all blocks are successful, the outputs are recombined into a bright amplified output beam. {\bf c)} Entanglement distillation with NLA. Alice has an EPR source and sends one of the two entangled beams to Bob through a lossy channel, inducing decoherence and degradation of shared entanglement. The loss in the channel is modelled with a beam splitter with reflectivity $\lambda$. Bob can purify the shared entanglement by utilising an NLA on his channel. }
\label{DistNLA}
\end{figure} 

The probabilistic NLA is depicted in Fig.~\ref{DistNLA}.a.  Whenever one and only one photon is detected on $d_1$ or $d_2$ the amplification is successful: an input state $\alpha \ket{0}+\beta \ket{1}$ is transformed (to within a normalisation factor) into $\alpha \ket{0} \pm g \beta \ket{1}$ where $g=\sqrt{\h/(1-\h)}$ is the amplitude gain related to the transmissivity $\h$ of the beam-splitter. In the following we omit the eventual $\pi$ phase shift on the $\ket{1}$ component as it can be compensated for. This 1-stage amplification process only acts on the vacuum and one photon components of the input state, and suppresses all the higher order photons. Small coherent states ($\ket{\alpha\ll1}$ have been experimentally amplified using a single stage NLA \cite{Fer10}. In order to amplify an arbitrary bright beam for which high order photon components cannot be neglected, one needs to first divide the input into $N\gg1$ modes. If $N$ is large enough, each mode will have a negligible component of $n>1$ photons and hence can be sent into the parallel 1-stage amplification stages depcited in Fig.~\ref{DistNLA}.b. When all these stages simultaneously succeed, the outputs are recombined resulting in an amplified bright beam. In the limit $N\to\infty$, a (globally) successful amplification leads to the coherent transformation $\ket{n}\to g^n\ket{n}$. This transformation can be used to probabilistically amplify coherent states without incurring a noise penalty: $\ket{\alpha}\sim\sum_n\frac{\alpha^n}{\sqrt{n!}}\ket{n} \to \sum_n \frac{\alpha^n}{\sqrt{n!}}g^n \ket{n}\sim\ket{g\alpha}$. 

We now consider a 2-mode EPR-entangled state that can be decomposed onto a Fock state basis as $\ket{\EPR^\chi_{AB}}\sim\sum_n \chi^n\ket{n_A, n_B}$ where $\chi$ represents the strength of the entanglement. Sending one of the modes into an NLA (Fig.~\ref{DistNLA}.c),  one can distil a more entangled state $\sum_n \chi^n g^n \ket{n_A, n_B}\sim\ket{\EPR^{g\chi}_{AB}}$. Moreover, it was shown that in principle a partially decohered EPR state caused by some loss $\l$ in the channel could also be conditionally purified as the effective loss of the line would be decreased by the NLA \cite{Ral09}. It is this property that we are interested in exploring in detail here.

The ideal transformation $\ket{n}\to g^n\ket{n}$ requires an infinite number of amplification stages. While the use of such a transformation to distil entanglement has already been explored theoretically \cite{Bla12}, the impracticality of a realistic physical implementation was overlooked. Here we analyse the performance of the NLA for a realistic case of a small number of stages, with criteria dealing with both entanglement and state purity. The simple question we ask is this: when does an ideal one or two-stage NLA surpass the usual entanglement and purity limitations due to line loss. By ideal NLA we assume that apart from the loss in the channel, all the optical components, photon detectors, and single photon sources are perfectly efficient. 

\section{Benchmarks}
\label{Benchmarks}

\begin{figure}[!!h]
\includegraphics[width=7cm]{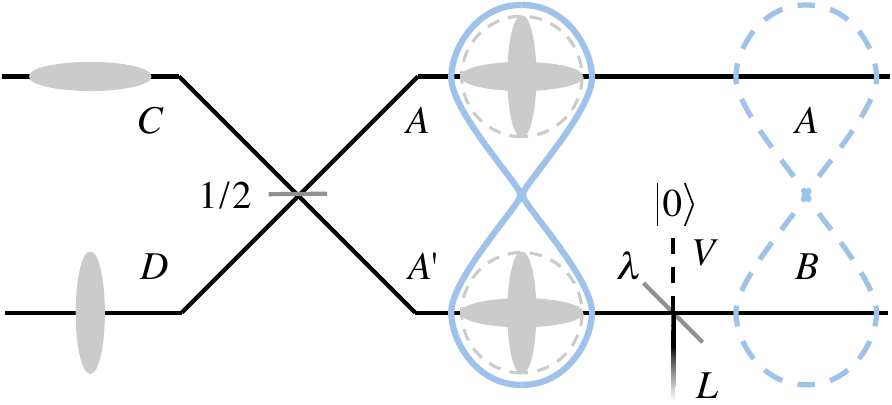}
\caption[No NLA]{A typical setup to distribute EPR entanglement. Two orthogonally squeezed vacua in modes $C$ and $D$ are mixed on a 50:50 beam splitter to create pure EPR entanglement in modes $A$ and $A'$. Mode $A'$ is then transmitted to $B$ trough a lossy channel. The loss is modelled by a beam splitter with reflectivity $\l$  splitting $A'$ into $B$ and the loss mode $L$, and a vacuum entering the second input of the beam splitter $V$. The loss results in a partly decohered EPR entanglement between modes $A$ and $B$.}
\label{NoNLA}
\end{figure} 

To set benchmarks for the usefulness of the NLA, we first study the best possible transmitted entanglement over a lossy channel. This situation corresponds to Fig.~\ref{DistNLA}.c when Bob doesn't use an NLA. 

\subsection{Best entanglement}

Our first benchmark is related to the improvement of entanglement due to the NLA. We choose the EPR criterion \cite{{mreidepr}} $\e$ as the measure of entanglement. The bipartite EPR criterion is directional, and comprises of two measures: $\e_{B|A}$ and $\e_{A|B}$ respectively defined as the products of the conditional variances of $B$ ($A$) with knowledge of $A$ ($B$):
\begin{eqnarray}
&\e_{B|A}=V^+_{B|A}V^-_{B|A},&\\
&V^\pm_{B|A}=\min_\gamma\!\left[ \Delta^2\!\left(\hat X^\pm_B-\gamma\hat X^\pm_A\right)\right],
\label{CondVar}
%&\e_{A|B}=V^+_{A|B}V^-_{A|B}&\nonumber\\
%&V^\pm_{A|B}=\min_\gamma\!\left(\hat X^\pm_A-\gamma\hat X^\pm_B\right)\nonumber\\
\end{eqnarray}
where $\Delta^2$ denotes the variance, $\hat X^+_M=\hat m+\hat m\dag$ and $\hat X^-_M=(\hat m-\hat m\dag)/i$ are the canonical quadrature operators of a general mode $M$, and $\hat m$ and $\hat m\dag$ are the annihilation and creation operators of mode $M$, respectively. $\e_{A|B}$ is defined symmetrically by swapping the direction by swapping the indices $A$ and $B$ in Eq.~\eqref{CondVar}. The conditional variances are normalised such that $\e_{A|B}$ and $\e_{B|A}$ denote entanglement when they are below 1, with 0 representing an unphysical limit of infinite entanglement strength.

To calculate $\e_{A|B}$ and $\e_{B|A}$ we first recall that the production of an EPR state can be modelled by mixing two equally but orthogonally squeezed vacua on a 50:50 beam splitter (Fig.~\ref{NoNLA}).
The squeezing operator is given by $\hat S_M(r)=\exp[r(\hat m^2-\left.\hat m\dag\right.^2)/2]$, where  $r$ is the squeezing parameter. Denoting $\hat{T}$ as the 50:50 beam splitter operation, we get
\begin{equation}
\hat{T} \hat S_C^r\hat S_D^{-r}\ket{0_{CD}} = \hat \s_{AA'}^r\ket{0_{AA'}},
\end{equation}
where we define the EPR operator on general modes $M$ and $N$ as
\begin{equation}
\hat \s^r_{MN} \equiv \exp[r(\hat m \hat n-\hat m\dag \hat n\dag)],
\end{equation}
which produces an EPR state when applied to the vacuum state as
\begin{eqnarray}
\hat \sigma_{MN}^r\ket{0_{MN}} &=& \sech(r) \sum_n \tanh(r)^n \ket{n_Mn_N}\nonumber\\
&=& \ket{\EPR_{MN}^{\tanh(r)}}.
\end{eqnarray}
Denoting the operation of a beam-splitter with reflectivity $\l$ as $\hat{T}_\l$, used in order to model loss on arm B, the final state can then be written as
\begin{eqnarray}
\ket{\Psi_{ABL}} &=& \hat{T}_\l \hat{T} \hat S_C^r\hat S_D^{-r}\ket{0_{CDV}}\nonumber \\ 
&=& \hat{T}_\l  \hat{T} \ket{r}_C\ket{-r}_D \ket{0_V},
\label{2modesqueezedstate}
\end{eqnarray}
where $\ket{r}_M=\hat S_M^r\ket{0_M}$ denotes the squeezed vacuum state in mode $M$ with squeezing parameter $r$. Using eq.(\ref{2modesqueezedstate}), and noting that eq.(\ref{CondVar}) is equivalent to
\begin{equation}
V^\pm_{B|A}=\Delta^2(\hat X^\pm_B)-\frac{C^\pm_{AB}}{\Delta^2(\hat X^\pm_A)}
\end{equation}
where $C^\pm_{AB}=\langle\hat X^\pm_A\hat X^\pm_B\rangle$ is the covariance,
we find
\begin{eqnarray}
%\e_{A|B} &=& \left(\frac{2 \sigma ^2+\lambda  \left(1-\sigma ^2\right)^2}{1+\sigma ^4-\lambda  \left(1-\sigma ^2\right)^2}\right)^2 \\
\e_{B|A} &=& \left[\l +(1-\l ) \sech(2r)\right]^2, \label{eqEntNoNLA}\\
\e_{A|B} &=& \frac{\e_{B|A}}{\left[1-\l\left(1-\sech(2r)\right)\right]^2},
\end{eqnarray}
which implies $\e_{B|A}<\e_{A|B}$. This means that in order to take maximum advantage of the entanglement, we should utilise the conditional knowledge Alice can obtain of Bob's state by measuring hers. We note that `sharing' the loss in the $A$-$B$ channel by having the EPR source in the middle ($i.e.$ both $A$ and $B$ experience half of the channel loss) leads to less entanglement. We thus define the remaining entanglement as $\e=\e_{B|A}$.

Fig.~\ref{EntNoNLA}.a shows the remaining entanglement $\e$ as a function of the loss $\l$ in the channel for various levels of input squeezing. Stronger levels of initial squeezing lead to stronger levels of entanglement, as can also be seen from Eq.~\eqref{eqEntNoNLA}. In the limit of infinite squeezing, one would obtain
\begin{equation}
\e_\infty\equiv\lim_{r\to\infty}\e=\l^2.
\label{benchmarkE}
\end{equation}
For a fixed level of loss $\l$, $\e_\infty$ sets a benchmark for the usefulness of the NLA: if by using an NLA one can obtain $\e_{\NLA}$ smaller than $\e_\infty$, this will provide unambiguous proof that the NLA allows one to achieve strengths of entanglement that would not be possible otherwise.

\subsection{Trade-off entanglement versus purity} 

\begin{figure}[!!h]
\begin{tabular}{rc}
%\begin{sideways}$\e_{A|B}$\end{sideways}&
%\begin{tabular}{c}
%\includegraphics[width=8cm]{EntNoNLAbad.pdf}
%\end{tabular} \\
{\bf a)}\\
$\e$&
\begin{tabular}{c}
\includegraphics[width=8cm]{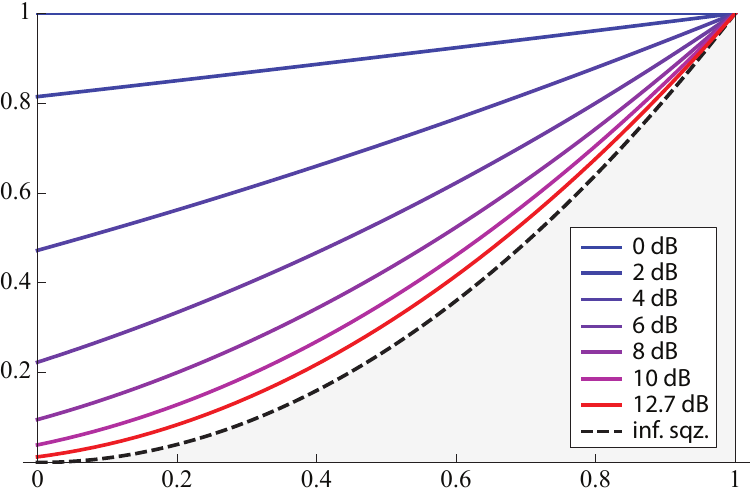}
\end{tabular} \\
{\bf b)}&$\l$\\
$p$&
\begin{tabular}{c}
\includegraphics[width=8cm]{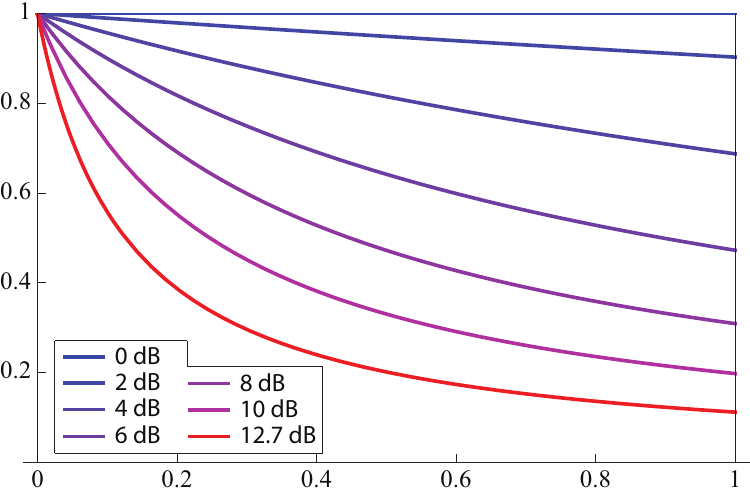}
\end{tabular} \\
&
$\l$
\end{tabular}
\caption[Entanglement without NLA]{{\bf a)} Level of remaining EPR entanglement  $\e$ and {\bf b)} purity $p$ as a function of the amount of loss $\l$ for various initial entanglement strength. The corresponding initial squeezing level varies from 0dB (blue) to 10dB (pink) by 2dB steps. The red curves correspond to the current world record of 12.7dB \cite{Ebe10}. The dashed black curves correspond to infinite squeezing, the gray area {\bf (a)} represents the domain inaccessible without NLA.}
\label{EntNoNLA}
\end{figure}

\begin{figure}[!h]
\begin{tabular}{rc}
$p$&
\begin{tabular}{c}
\includegraphics[width=7.65cm]{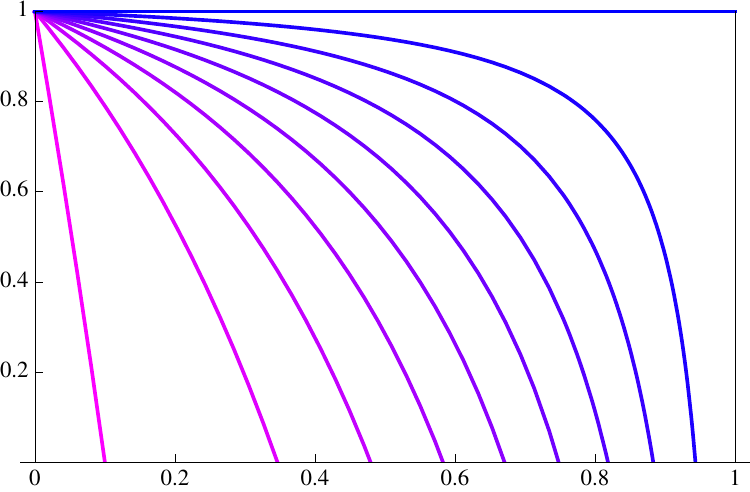}
\end{tabular} \\
&
$\l$
\end{tabular}
\caption[Budget without NLA]{Purity-entanglement trade-off curves. The final purity $p$ is plotted against the loss level $\l$ for various required final entanglement $\e_{B|A}$ ranging from 1 (blue) to 0.01 (pink) by steps of 0.11.}
\label{Budget}
\end{figure}

We select the final purity of the quantum state as our second benchmark, as it is critical in some quantum information protocols such as quantum key distribution~\cite{Gro03,Sym07} and quantum teleportation \cite{Vai94,Fur98,Bow03}.
The purity $p$ of a mixed state represented by a density matrix $\hat \r$, is defined as $\Tr(\hat \r^2)$, and is easily computable from the square integral of the Wigner function. In the absence of an NLA, and for a loss $\l$ on one of the modes, we find the purity of the two-mode entangled state to be
\begin{equation}
p=\frac{1}{1+\lambda( \cosh(2r) -1)}.
\label{pNoNLA}
\end{equation}
 Eliminating $r$ in Eqs.~\eqref{eqEntNoNLA} and~\eqref{pNoNLA} leads to
\begin{equation}
p=\frac{1-\frac{\l }{\sqrt{\e }}}{1-\l },
\label{benchmarkP}
\end{equation}
which allows us to obtain the purity given a desired level of entanglement $\e$, after a loss $\l$. This implies that a higher level of output entanglement strength comes at the expense of a smaller output purity as clearly shown in Fig.~\ref{EntNoNLA}.b, which in turn implies a necessary trade-off between high entanglement and high purity. Fig.~\ref{Budget} displays these `trade-off' curves $p(\e,\l)$ for various $\e$ values. These curves also set a benchmark for the efficacy of the NLA: for a given amount of loss, if one can obtain values above the curve $p(\e,\l)$ with an NLA, this will provide evidence that the NLA allows for better purity than is otherwise accessible.

\section{Single stage NLA}
\label{1stage}

\begin{figure}[h]
\includegraphics[width=7cm]{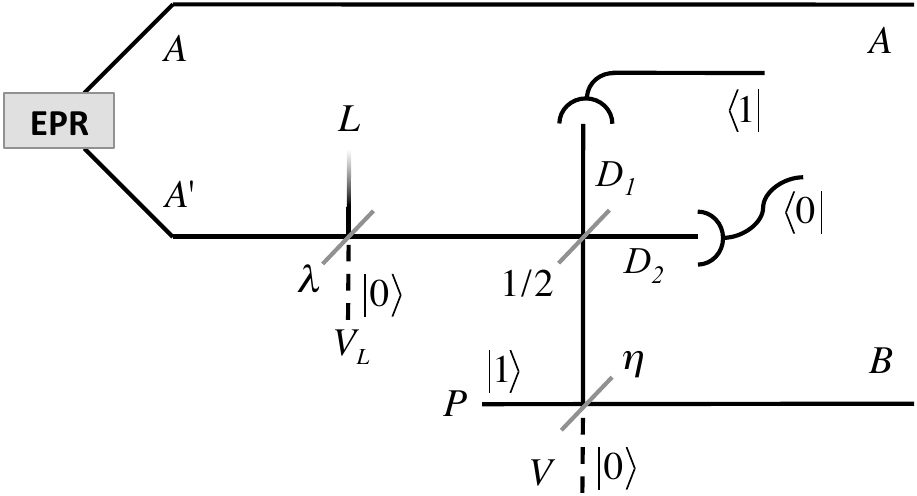}
\caption[No NLA]{ Schematics of the distillation of a partly decohered EPR state by use of a one stage NLA after the lossy channel. $A$ and $A'$ are the pure entangled beams. The lossy channel is modelled by a beam splitter with reflectivity $\l$ towards a loss mode $L$. A vacuum enters the second port $V_L$ of the beam splitter. $\h$ and $1/2$ represent the two beam splitters of the NLA. $P$, $V$, $D_1$ and $D_2$ respectively are the auxiliary single photon and vacuum inputs and detection modes of the NLA. Finally, $B$ is the output of the NLA. }
\label{1SNLA}
\end{figure} 

In this section we consider the case of a single stage NLA, and address the regimes for which it can beat the two benchmarks defined in Eqs.~\eqref{benchmarkE} and~\eqref{benchmarkP} established in the previous section. The layout for such a distillation process is depicted in Fig.~\ref{1SNLA}. In part A we derive the state produced by a successful distillation. In part B we show that in the limit of infinitely low success probability, the benchmarks can be beaten. Finally, in parts C and D we show how the benchmarks can be beaten with realistic success rates.

\subsection{Derivation of the distilled quantum state}

 The state produced by this setup when the amplification is successful is defined as
\begin{eqnarray}
\ket{\Psi_{ABL}} &=& \bra{1_{D_1}}\!\bra{0_{ D_2}}\hat{\mathcal{T}} \ket{\EPR_{AA'}}\ket{1_P}\ket{0_V}\ket{0_{V_L}}\nonumber\\
&=&\braopket{0_{D_1 D_2}}{\hat d_1 \hat{\mathcal{T}} \hat p\dag \hat \s^r_{AA'} }{0_{AA'PV_LV}}
\end{eqnarray}
where $\hat{\mathcal{T}}$ denotes the operation of the three beam splitters. We now reduce this expression to a simpler one that will allow for an easy calculation of the purity and entanglement of the amplified state. To do so, we first notice that the EPR operator $\hat{\s}_{AB}^r$ can be easily expressed in a `normally ordered' form when applied to the vacuum:
\begin{equation}
\hat \s_{AA'}^r\ket{0_{AA'}}=\sech r \exp [\tanh(r)  \hat a\dag\hat a'\vphantom{a}\dag]\ket{0_{AA'}}.
\end{equation}
We then commute $\mathcal T$ to the right, transforming the argument of the exponential into an expression now containing products of creation operators in mode $A$ and modes $L$, $D_1$, $D_2$ with various coefficients, one of them, for example, being $\sqrt{(1-\l)/2}\tanh(r)\hat a\dag \hat d_1\dag$. The annihilation operator $\hat d_1$ is then commuted to the right using standard quantum algebra, the commutation with the exponential giving rise to an additionnal term $\sqrt{(1-\l)/2}\tanh(r)\hat a\dag $ (outside the exponential). We are now left with the exponential applied to the vacuum in modes $D_1$ and $D_2$ and projected back onto the vacuum in modes $D_1$ and $D_2$. This exponential  containing creation operators only, we can formally discard  all the terms containing  $\hat d_1\dag$ or $\hat d_2\dag$ in its argument, reducing the expression to $\exp[\sqrt\l\tanh(r) \hat a\dag \hat l\dag]$ which can be identified with $\cosh(\r)\hat \s^\r_{AL}$ where $\r$ is defined by $\tanh(\r) = \sqrt \l \tanh(r)$. We finally obtain
\begin{eqnarray}
\ket{\Psi_{ABL}}& =&\frac{\cosh(\r)}{\sqrt2\cosh(r)}\left(\sqrt{1-\h}\right.\\
&&+\left.\sqrt{\h(1-\l)}\tanh(r)\hat a\dag \hat b\dag\right)\hat\s^\r_{AB}\ket{0_{ABL}}\nonumber
\end{eqnarray}
which may be more conveniently rewritten as (ignoring normalisation)
\begin{equation}
\ket{\Psi_{ABL}}\sim(1+\k \; \hat a\dag \hat b\dag)\hat \sigma^\rho_{AL}\ket{0_{ABL}}
\label{NLA1stage}
\end{equation}
where \begin{equation}
\k=g\sqrt{1-\l} \tanh(r).
\label{kappa}
\end{equation}

\subsection{Beating the benchmarks}

The term $1+\k \hat a\dag \hat b\dag$ in Eq.~\eqref{NLA1stage} corresponds to the production of the first order truncated EPR state $\ket{0_A0_B}+\k\ket{1_A1_B}$. The term $\hat\s^\r_{AL}$ in the same equation corresponds to the production of an EPR state between mode $A$ and the loss mode $L$. Entanglement between the system $A/B$ and the environment is generated, inducing decoherence. The entanglement strength between $A$ and $L$ is given by $\tanh(\r) = \sqrt \l \tanh(r)$. For any non-zero loss $\l$, it is possible to suppress this decoherence by taking the limit $r\to 0$. Note that this is also the case without an NLA: with no initial entanglement ($i.e.$ using the vacuum as the initial state), there is no decoherence. Obviously, there is also no entanglement at the output. However, using an NLA it is possible to increase the gain $g\to\infty$ as we decrease $r\to0$, keeping $\k$ constant at a non-zero value. This allows one to distil an arbitrarily pure, although truncated to the first order, EPR state between modes $A$ and $B$. 
 
The maximum EPR entanglement one can obtain from a state of the form $\ket{00}+\k\ket{11}$ is when $\k
% = \sqrt{\left(2 \sqrt{6}-3\right)/15}
\simeq0.36$ and is $\e
%=40-16 \sqrt{6}
\simeq0.81$. 
%The ability of obtaining arbitrarily pure states with down to $\e\simeq0.81$ EPR entanglement clearly breaches the two benchmarks (\ref{benchmarkE}) and (\ref{benchmarkP}).
Approaching this value requires $r\to0$ and $g\to\infty$, which means that the probability of success $\Pi$ of the NLA goes to 0:
 \begin{eqnarray}
\Pi&=&2\braket{\Psi_{ABL}}{\Psi_{ABL}}\nonumber\\
&=&\frac{ 1-\eta +(\eta -\lambda ) \tanh(r)^2}{ \left(1-\lambda  \tanh(r)^2\right)^2\cosh(r)^2} \label{ProbSuccess}\\
\nonumber\\
&\underset{\begin{tiny}\begin{tabular}{c}$r\to0$\\$\h\to1$\end{tabular}\end{tiny}}{\longrightarrow}&0.\nonumber
\end{eqnarray}
The factor of 2 in Eq.~\eqref{ProbSuccess} accounts for the fact that $\braket{\Psi_{ABL}}{\Psi_{ABL}}$ represents the probability of detecting $\bra{1_{D_1}0_{D_2}}$ only. The probability of detecting $\bra{0_{D_1}1_{D_2}}$ is identical. 

The trade-off between entanglement level and purity has been shifted to a trade-off between the probability of success and entanglement/purity. In the limit $\Pi\to0$ the two benchmarks established in section \ref{Benchmarks} can be beaten: for any value of $\l$, one can distil an arbitrarily pure state down to $\e=0.81$. The relevant question is now: to what extent can we beat the benchmarks with reasonable non-zero probabilities of success?

\subsection{Beating the entanglement benchmark with finite probability of success}

To estimate $\e$ (we note that $\e=\e_{B|A}$ is also the best choice with an NLA), we need to calculate terms such as $\langle\Psi_{ABL}|(\hat X_A^+)^2|\Psi_{ABL}\rangle$ which are of the form $\langle0_{ABL}|\hat\s_{\!AL}^\r\vphantom{)}\!\dag f(\hat a,\hat a^{\dagger},\hat b,\hat b^{\dagger})\hat\s_{\!AL}^\r|0_{ABL}\rangle$, where $f$ is a polynomial function of the creation and annihilliation operators in modes $A$ and $B$. 
The terms $\hat a^{\dagger}$, $\hat b^{\dagger}$, and their respective Hermitian conjugates can be commuted to the right using the transformation 
\begin{equation}
\hat m \hat \s_{MN}^r = \cosh(r) \hat \s_{MN}^r \hat m+\sinh(r)\hat\s_{MN}^r\hat n,
\end{equation}
which leads (noting $\hat\s_{\!AL}^\r\vphantom{)}\hat\s_{\!AL}^\r\vphantom{)}\!\dag=1$) to an expression of the form $\langle0_{ABL}|h(\hat a,\hat a^{\dagger},\hat b,\hat b^{\dagger})|0_{ABL}\rangle$, where $h$ is a new polynomial function. This last expression is easily computable using standard quantum algebra.

%Such a calculation leads to a expression of $\e$ depending on $r$, $\h$ and $\l$. 
Such a calculation leads to an expression $\e(r,\l,\h)$ which may be transformed using Eq.~\eqref{ProbSuccess} into a new expression $\e(r,\l,\Pi)$. For a given loss in the channel $\l$ and a non-zero probability of success $\Pi$, the initial entanglement level $r$, as well as the corresponding gain in the NLA required to maintain $\Pi$ at the chosen level, can be optimised to produce the best output entanglement possible after distillation, given by $\e_{\text{opt}}=\min_r \e(r,\l,\Pi)$:
\begin{widetext}
\begin{equation}
\begin{split}
%\e=&\min_r\left[\left(1-2 \l +\frac{4}{\Pi }+2 (-1+\l ) \cosh(2 r)+\frac{8}{-(1+\l ) \Pi +(-1+\l ) \Pi  \cosh(2 r)}+ \right. \right. \\ 
%& \left. \left. \left(8 \left(1+\Pi  \cosh(r)^4+\l  \sinh(r)^2+\l ^2 \Pi  \sinh(r)^4-\cosh(r)^2 \left(1+2 \l  \Pi  \sinh(r)^2\right)\right) \left(1+\l  \tanh(r)^2\right)\right)/\left(-4+3 \Pi +\l  (4+2 \Pi +3 \l  \Pi )+2 (-1+\l ) (-2+(-1+\l ) \Pi ) \cosh(2 r)-(-1+\l )^2 \Pi  \cosh(4 r)-4 \l ^2 \Pi  \sech(r)^2\right)\right)^2\right]
\e_{opt} =&\min_r\left( 1-2 \l +\frac{4}{\Pi }-2 (1-\l ) \cosh(2 r)-\frac{8}{(1+\l ) \Pi +(1-\l ) \Pi  \cosh(2 r)}+  \right. \\
& \left. \frac{8 \left(1+\Pi  \cosh(r)^4+\l  \sinh(r)^2+\l ^2 \Pi  \sinh(r)^4-\cosh(r)^2 \left(1+2 \l  \Pi  \sinh(r)^2\right)\right) \left(1+\l  \tanh(r)^2\right)}{-4\!+\!3 \Pi\! +\!\l  (4\!+\!2 \Pi \!+\!3 \l  \Pi )\!+\!2 (1\!-\!\l ) (2\!+\!(1\!-\!\l ) \Pi ) \cosh(2 r)\!-\!(1\!-\!\l )^2 \Pi  \cosh(4 r)\!-\!4 \l ^2 \Pi  \sech(r)^2}\right)^2.
\end{split}
\end{equation}
\end{widetext}

\begin{figure}[ht]
\begin{tabular}{rc}
{\bf a)}\\
$\e$&
\begin{tabular}{c}
\includegraphics[width=7.5cm]{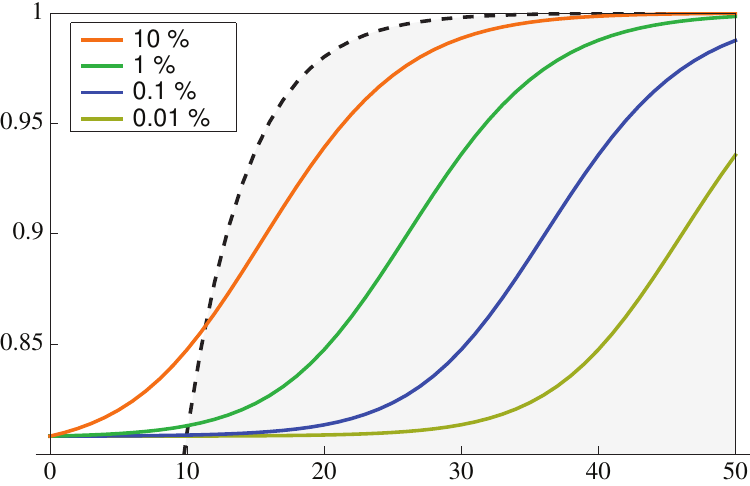}
\end{tabular} \\
{\bf b)}&
loss (dB)\\
$p$&
\begin{tabular}{c}
\hspace{0.1cm}
\includegraphics[width=7.5cm]{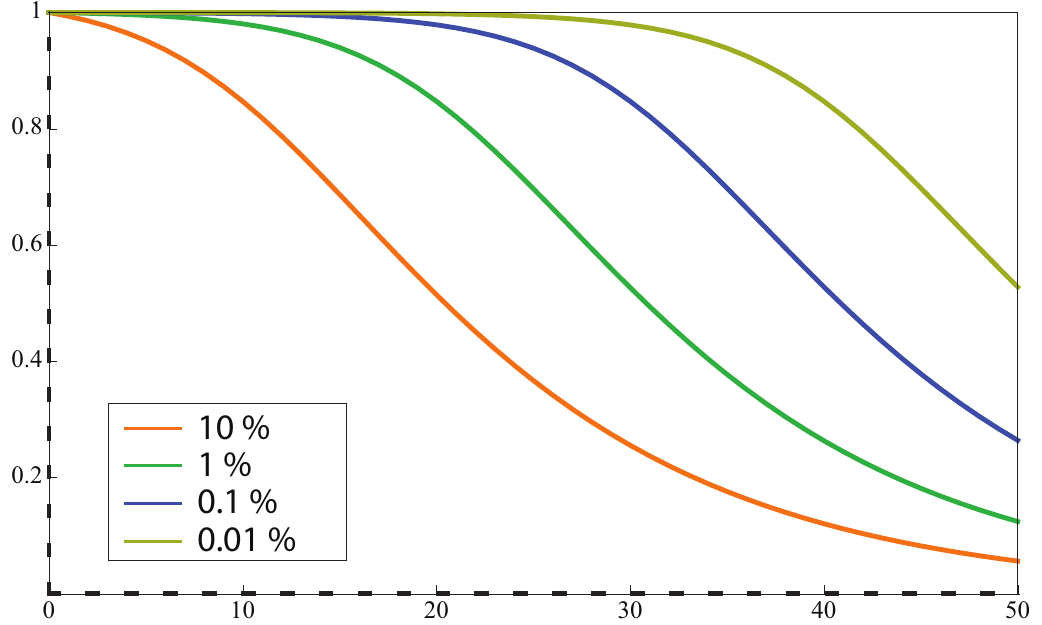}
\end{tabular} \\
&
loss (dB)
\end{tabular}
\caption[Results 1 stage]{{\bf a)} Final entanglement $\e$ and {\bf b)} purity $p$ after distillation optimised for entanglement strength with a 1-stage NLA against the loss in logarithmic scale $\l=1-10^{-dB/10}$, for various probabilities of success: $\Pi=10\%$ (orange), 1\% (green), 0.1\% (blue) and 0.01\% (yellow). The black dashed curve reminds the best entanglement possible without NLA {\bf (a)}, which implies infinitely small purity {\bf (b)}, and the grey area the domain inaccessible without NLA {\bf (a)}. 
%The red dashed curve (b) corresponds to the purity without NLA and final entanglement set to 0.85. The fattened parts evidence regimes with enhancement of both purity and entanglement strength.
}
\label{Results1S}
\end{figure}

Fig.~\ref{Results1S}.a shows the results of such an optimisation for success probabilities ranging from $\Pi=10\%$ to 0.01\%. On the one hand, it appears clearly that the distillation is  beneficial for a channel with more than $\sim10$dB loss. The figure also shows that every additional 10dB of loss in the channel can be compensated by giving up an additional 10dB in the success rate $\Pi$ of the distillation. On the other hand, for a given amount of loss, the distilled entanglement saturates at $\sim0.81$ when $\Pi\to0$. This means that the entanglement level is no longer limited by the purity of the output state but only by the absence of higher photon number terms. To further improve the entanglement level, one would need to use a multi-stage NLA.

To verify this last statement the corresponding purity of the state is computed, and plotted in Fig.~\ref{Results1S}.b). The purity $p$ of the state is given by $\Tr\![\hat \r_{\!A\!B}^2]$ where $\hat \r_{\!A\!B}=\Tr_{\!\!L}[\ket{\Psi_{\!A\!B\!L}}\!\!\bra{\Psi_{\!A\!B\!L}}]$. The calculation of $p$ leads to an expression $p(r,\l,\h)$ which can be transformed, as before, into $p(r,\l,\Pi)$ by using Eq.~\eqref{ProbSuccess}:
\begin{widetext}
\begin{equation}
\begin{split}
p(r,\l,\Pi)=&
\left[\left(1-\l  \tanh ^2(r)\right) \left( \vphantom{\left(\l \tanh ^2(r)+1\right)^2}2 \sinh ^2(r) \left(-2
   \l +\l ^3 \tanh ^4(r)+\l  \tanh ^2(r)+1\right) +\right.\right.\\
   &\left.\left. \l ^2
   \Pi ^2 \sinh ^8(r) \left(5 \l ^2-8 \l +\l ^2 \left(2
   \l ^2-2 \l +1\right) \tanh ^4(r)-2 \l  \left(3 \l
   ^2-3 \l +1\right) \tanh ^2(r)+2\right)+\right.\right.\\
   &\left.\left.\cosh ^4(r) \left(2 \Pi
   -\left(4 \l ^2-2 \l -1\right) \Pi ^2 \sinh ^4(r)-2 (\l -2)
   \Pi  \sinh ^2(r)+1\right)+\right.\right.\\
   &\left.\left.2 \cosh ^2(r) \left(\l  \left(\l ^2+2
   \l -1\right) \Pi ^2 \sinh ^6(r)+\left(3 \l ^2-1\right) \Pi 
   \sinh ^4(r)+(\l -1) (\Pi +1) \sinh ^2(r)-1\right)+\right.\right.\\
   &\left.\left.2 \l  \Pi 
   \sinh ^6(r) \left(\l ^2-4 \l +\l ^2 \left(2 \l ^2-2
   \l +1\right) \tanh ^4(r)-(1-2 \l )^2 \l  \tanh
   ^2(r)+1\right)+\right.\right.\\
   &\left.\left.\sinh ^4(r) \left(-\l ^2 (4 \Pi +1)-2 \l
   +\l ^2 \left(2 \l ^2 (\Pi +1)-2 \l +1\right) \tanh ^4(r)-2
   (\l -1) \l ^2 (\Pi +1) \tanh ^2(r)+1\right)+\right.\right.\\
   &\left.\left.\left(\l 
   \tanh ^2(r)+1\right)^2+\Pi ^2 \cosh ^8(r)-2 \Pi  \cosh ^6(r) \left(\Pi 
   \sinh ^2(r)+1\right)\right)\right] \\
   & /\left[\left(\l  \tanh ^2(r)+1\right)^3
   \left(\l  \Pi  \sinh ^4(r) \left(-\l +(\l -1) \l 
   \tanh ^2(r)+2\right)-\cosh ^2(r) \left((\l +1) \Pi  \sinh
   ^2(r)+1\right)+\right.\right.\\
   &\left.\left.\sinh ^2(r) \left((\l -1) \l  \tanh
   ^2(r)+1\right)+\Pi  \cosh ^4(r)+1\right)^2\right].
\end{split}
\label{eq:purity_long}
\end{equation}
\end{widetext}

If we use $r_{\text{opt}}(\l,\Pi)$, defined by $\e_{\text{opt}}=\e(r_{\text{opt}},\l,\Pi)$, in Eq.~\eqref{eq:purity_long} we may obtain the purity corresponding to the optimal entanglement.

Fig.~\ref{Results1S}.b exhibits the aforementioned correlation between the drop of purity and loss of entanglement alluded to in Fig.~\ref{Results1S}.a. Also evident is that even with reasonable success rates, and parameters optimised for strongest levels of entanglement, the final state can be very pure over a wide range of loss.

% can be calculated and expressed as $p(r,\l,\h)$ or, equivalently with (\ref{ProbSuccess}), $p(r,\l,\Pi)$ which can then be fed with the result $r(\l,\Pi)$ found by the optimisation $\e=\min_r \e(r,\l,\Pi)$.. 
%Using the Fock state basis, one finds
%\begin{eqnarray}
%p &= &\frac{\sech(\r)^4}{\left(1+\k ^2\right)^2}\sum_n \left[\tanh(\r)^{2n}\left(1+\k ^2(n+1)\right)\right]^2 \nonumber \\
%%&=&\frac{s^4 \left(-2 \k ^4-s^2 \left(-2+s^2\right) \left(-2 \k ^2+\k ^4-2 s^2+s^4\right)\right)}{\left(1+\k ^2\right)^2 \left(-1+\tanh(\r)^4\right)^3}
%&=&\frac{\left(1+\k ^2\right)^2+\left(-2-2 \k ^2+\k ^4\right) \l ^2\tau^4+\l ^4\tau^8}{\left(1+\k ^2\right)^2 \left(1-\l \tau^2\right) \left(1+\l \tau^2\right)^3}
%\end{eqnarray}
%where $\tau=\tanh[r]$. 
%Injecting (\ref{kappa}), we find an expression $p(r,\l,\h)$ or, equivalently with (\ref{ProbSuccess}), $p(r,\l,\Pi)$ which can then be fed with the result $r(\l,\Pi)$ found by the optimisation $\e=\min_r \e(r,\l,\Pi)$.

\subsection{Beating the purity benchmark at finite probability of success}

 \begin{figure}[ht]
\begin{tabular}{rc}
$p$&
\begin{tabular}{c}
\hspace{0.1cm}
\includegraphics[width=8cm]{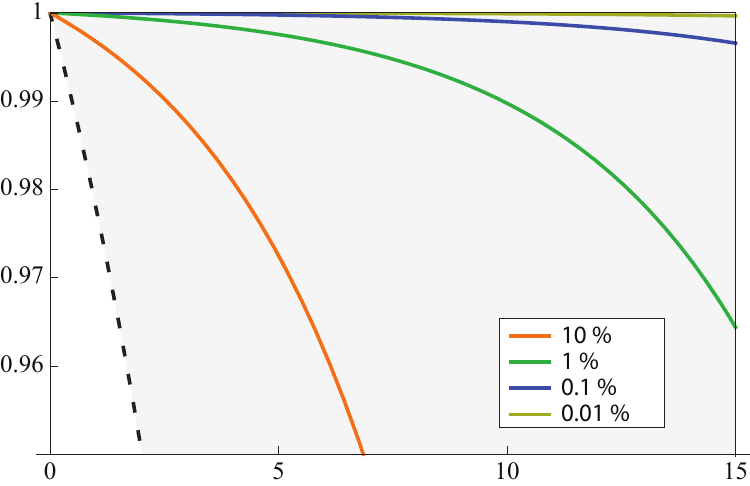}
\end{tabular} \\
&
loss (dB)
\end{tabular}
\caption[Results 1 stage]{Purity of a $\e=0.85$ entangled state as a function of loss $\l=1-10^{-dB/10}$ for probability of success: $\pi=10\%$ (orange), 1\% (green), 0.1\% (blue) and 0.01\% (yellow). The black dashed curve reminds the corresponding benchmark curve when no NLA is used, and hence the gray area represents the domain inaccessible without NLA.}
\label{Pur1S}
\end{figure}

While a 1-stage NLA does not appear to be necessarily beneficial for distilling stronger entanglement when the loss is less than $\sim$10dB, it can still be used to distil purer entanglement with a relatively high success rate. Let us recall that studying purity is relevant only for a fixed level of final entanglement (as perfect purity is always achieved with vacuum). Hence we transform our expression $p(r,\l,\Pi)$ into $p(\e,\l,\Pi)$ by inverting $\e(r,\l,\Pi) \rightarrow r(\e,\l,\Pi)$. 
Fig.~\ref{Pur1S} shows how a fixed level of entanglement ($\e=0.85$) can be purified beyond the no-NLA benchmark, as the success rate increases. Note that with this criterion, using an NLA is beneficial for any value of the loss level $\l$.

% Without NLA, this implies a compromise between entanglement strength and purity, whereas with NLA there is no limit on how pure the state can be and the compromise lies only with how low the probability of success can be. Fig.\ref{Results1S}.b shows that even with reasonable success rates, the final state can be very pure over a wide range of loss. Here again we see that every additional 10dB of loss in the channel can be compensated by giving up an additional 10dB in the success rate. More importantly, the fattened parts of the curves conservatively evidence regimes where both the purity and  entanglement are better than without NLA. While NLA does not appear necessary beneficial for distilling stronger entanglement when the loss is less than $\sim$10dB, it can be used to distill purer entanglement even when the loss is smaller than $\sim$10dB.

\section{Dual stage NLA}

In the previous section we showed that for reasonable success rates, many situations can be found where using a 1-stage NLA gives strong enhancement in terms of entanglement strength and/or purity. However, because a 1-stage NLA cannot give access to more that one photon in mode $B$, the distilled entanglement can never reach bellow $\e=0.81$. To go beyond this limit, more stages have to be implemented.

Following the same procedure as described in section~\ref{1stage}, we find that the output state for a 2-stage NLA is 
%\begin{equation}
%\ket{\Psi_{ABL}}\sim\left[1+\k a\dag b\dag+\frac12\left(\k a\dag b\dag\right)^2\right]\sigma^\rho_{AL}\ket{0_{ABL}}.
%\label{NLA2stage}
%\end{equation}
%\begin{equation}
%\ket{\Psi_{ABL}}\sim\left[1+\k \; \hat a\dag \hat b\dag+\frac{\k^2}4 \left.\hat a\dag\right.^2 \left.\hat b\dag\right.^2\right]\hat \sigma^\rho_{AL}\ket{0_{ABL}}.
%\label{NLA2stage}
%\end{equation}
\begin{equation}
\ket{\Psi_{ABL}}=\xi\left[1+\k \; \hat a\dag \hat b\dag+\frac{\k^2}4 \left.\hat a\dag\right.^2 \left.\hat b\dag\right.^2\right]\hat \sigma^\rho_{AL}\ket{0_{ABL}},
\label{NLA2stage}
\end{equation}
with
\begin{equation}
\xi =\frac{\cosh(\r)}{\cosh(r)}\;\frac{1-\h}{2}.
\end{equation}
We note that $\xi$ is not the norm of $\ket{\Psi_{ABL}}$.
The term $\frac{\k^2}4 \left.\hat a\dag\right.^2 \left.\hat b\dag\right.^2$ now gives access to the two photon component, thus allowing 
%corresponds \red{\sout{to the production of an EPR states truncated to the second order: $\ket{00}+\k\ket{11}+\k^2\ket{22}$.} No! The 2 photon term is $\k^2/2 \ket{22}$ } The second order terms allow access to 
a stronger entanglement. The maximum entanglement for a state $\ket{00}+\k\ket{11}+\frac{\k^2}{2}\ket{22}$ is obtained for $\k\simeq0.59$ and is $\e\simeq0.57$. 

\begin{figure}[t]
\begin{tabular}{rc}
a)\\
$\e$&
\begin{tabular}{c}
\includegraphics[width=8cm]{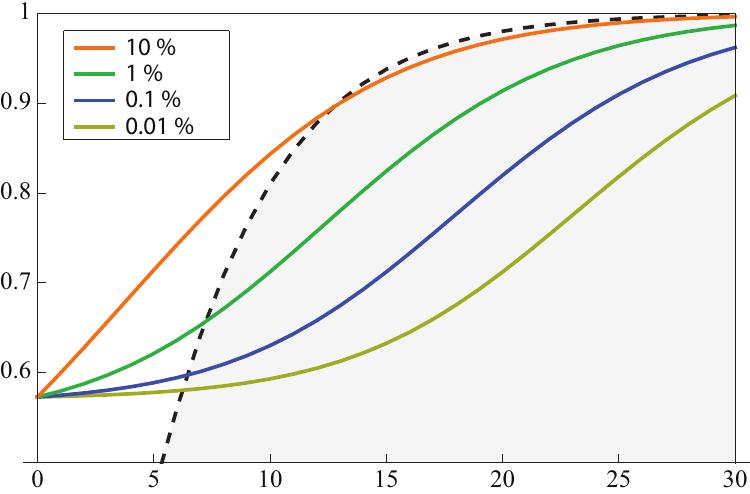}
\end{tabular} \\
b)&loss (dB)\\
$p$&
\begin{tabular}{c}
\hspace{0.1cm}%\includegraphics[width=8cm]{Pur2S.pdf}
\includegraphics[width=8cm]{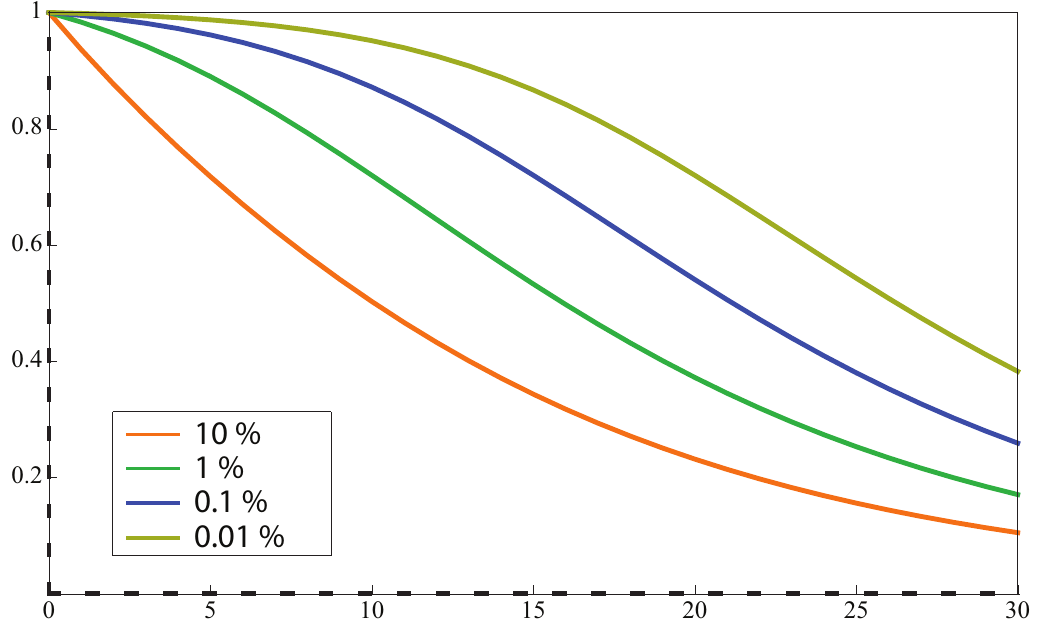}
\end{tabular} \\
&
loss (dB)
\end{tabular}
\caption[Results 1 stage]{{\bf a)} Final entanglement $\e$ and {\bf b)} purity $p$ after distillation optimised for entanglement strength with a 2-stage NLA against the loss $\l=1-10^{-dB/10}$, for probability of success: $\pi=10\%$ (orange), 1\% (green), 0.1\% (blue) and 0.01\% (yellow). The black dashed curve reminds the best entanglement possible without NLA {\bf (a)}, which implies infinitely small purity {\bf (b)}, and the grey area the domain inaccessible without NLA {\bf (a)}.
%The red dashed curve (b) corresponds to the purity without NLA and final entanglement set to 0.65.
}
\label{Results2S}
\end{figure}

\begin{figure}[!!h]
\begin{tabular}{rc}
$p$&
\begin{tabular}{c}
\hspace{0.1cm}%\includegraphics[width=8cm]{BestPur2s.pdf}
\includegraphics[width=8cm]{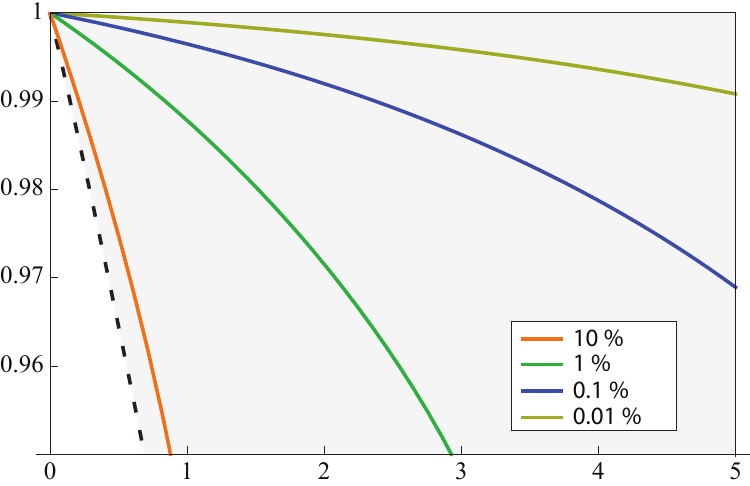}
\end{tabular} \\
&
loss (dB)
\end{tabular}
\caption[Results 1 stage]{Purity of a $\e=0.6$ entangled state as a function of loss $\l=1-10^{-dB/10}$ for probability of success: $\pi=10\%$ (orange), 1\% (green), 0.1\% (blue) and 0.01\% (yellow). The black dashed curve reminds the corresponding benchmark curve when no NLA is used, and hence the gray area represents the domain inaccessible without NLA.}\label{Pur2S}
\end{figure}

Following the analysis in section~\ref{1stage} we find the best entanglement $\e$ achievable with loss $\l$ and fixed success rate $\Pi$ by optimising values of $r$ and $\h$, as depicted in Fig.~\ref{Results2S}. First, we remark that a 2-stage NLA can be beneficial starting from $\sim$6dB of loss compared to the $\sim$10dB threshold imposed in the single stage case. Second, we observe that every additional 10dB of loss has to be compensated by giving up an additional 20dB in the success rate. This is due to being penalised 10dB with respect to success rates for each of the two individual stages. Fig.~\ref{Pur2S} shows how a stronger fixed level of entanglement of $\e=0.6$ can be purified beyond the non-NLA benchmark as the success rate increases.

\begin{figure}[!!h]
\begin{tabular}{rc}
a)\\
$\e$&
\begin{tabular}{c}
\includegraphics[width=8cm]{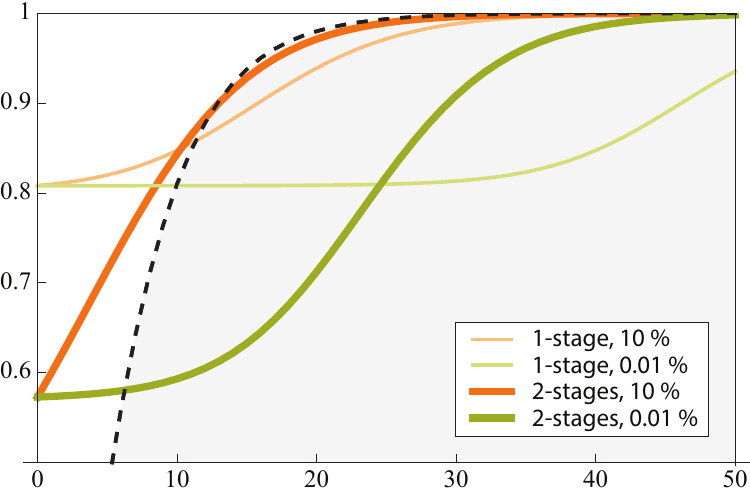}
\end{tabular} \\
b)&
loss (dB)\\
p&
\begin{tabular}{c}
\includegraphics[width=8cm]{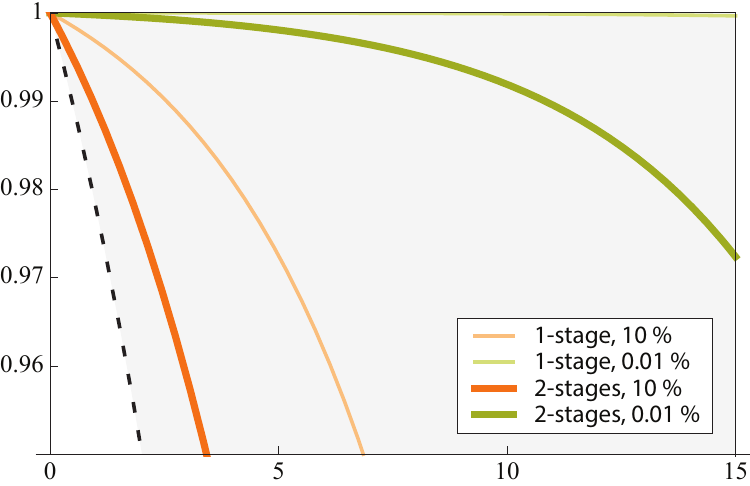}
\end{tabular} \\
&
loss (dB)
\end{tabular}
\caption[Comparison 1-2 stage]{Comparison of the 1-stage (light/thin curves) and 2-stages (fat curves) NLA operating at 10\% (orange) and 0.01\% (yellow) probability of success. {\bf a)} Distillation optimised for maximum entanglement. The dashed black curves correspond to entanglement without NLA with infinite squeezing. {\bf b)} Distillation of purer states with a given entanglement of $\e=0.85$ The dashed curve correspond to purity without use of NLA. The gray areas correspond to the domains inaccessible without NLA.}
\label{Comp12}
\end{figure}

It is interesting to compare the performances of the 1-stage NLA and the 2-stage NLA in regard to distillation when operating at the same success rate. Fig.~\ref{Comp12} summarises the results of the both distillation processes for two regimes: high success rate (10\%) and low success rate (0.01\%). Fig.~\ref{Comp12}.a shows that a 2-stage distillation is never beneficial if the aim is to obtain the strongest entanglement combined with a high success rate. In the regime where the 2-stage NLA surpasses the 1-stage NLA (loss$<\sim$10dB), it is in fact better to  utilise the strongest entanglement possible, importantly with no distillation. This is because for high success rates, the entanglement value is not limited by the absence of larger photon numbers but by the purity of the states. In the low success rate regime, the 2-stage NLA does not necessarily surpass the 1-stage NLA. For example, at the success rate of 0.01\%, the 2-stage NLA surpasses the 1-stage NLA only for losses smaller than $\sim$25dB. 

Finally Fig.~\ref{Comp12}.b illustrates that a 2-stages distillation is not beneficial over a 1-stage if the aim is to obtain a better purity, as long as the entanglement required is achievable with a 1-stage NLA, i.e. for $\e>0.81$.
To maintain the success rate when the loss increases, the incoming photon flux in each stage has to increase, whilst lowering the gain. This implies that the initial entanglement strength has to increase faster for a given loss when the NLA has 2 stages compared to when it has only 1 stage. 
%as evidenced in Fig.\ref{InputEnt}. 
Consequently the effect of decoherence appears faster with the 2-stages NLA.

%\begin{figure}[!!h]
%\begin{tabular}{rc}
%a)\\
%\begin{sideways}$\e_{in}$\end{sideways}&
%\begin{tabular}{c}
%\includegraphics[width=8cm]{InputEnt1S.pdf}
%\end{tabular} \\
%b)&loss (dB)\\
%\begin{sideways}$\e_{in}$\end{sideways}&
%\begin{tabular}{c}
%\hspace{0.1cm}\includegraphics[width=8cm]{InputEnt2S.pdf}
%\end{tabular} \\
%&
%loss (dB)
%\end{tabular}
%\caption[Results 1 stage]{Input entanglement $\e_{in}=4e^{2r}/\left(1+e^{4r}\right)^2$ for {\bf a)} the 1-stage and {\bf b)} 2-stages against the loss $\l=1-10^{-dB/10}$, for probability of success: $\pi=10\%$ (orange), 1\% (green), 0.1\% (blue) and 0.01\% (yellow).}
%\label{InputEnt}
%\end{figure}

%\red{
%I don't really want to talk about that as it is linked to experimental imperfections which I decided to overlook. 
%Or I can try to pull out some results that include realistic imperfections...
%Fig.\ref{InputEnt} also shows that the input entanglement doesn't have ridiculous values.
%Fig.\red{blah} shows the $\h$ values. }

\section{Multi stage NLA}

In the case of an arbitrary number of stages $N$, Eqns.~\eqref{NLA1stage} and~\eqref{NLA2stage} pertaining to the distilled state can be generalised to
\begin{equation}
\ket{\Psi_{\text{N-stage}}}\sim\left(1+\frac\kappa N \hat a\dag\hat b\dag\right)^N\hat\s_{AL}^\r\ket{0_{ABL}}.
\label{eq:multiNLA}
\end{equation}
In the limit of perfect distillation, $r\to0$ and $g\to\infty$ at constant $\k$ (also implying $\Pi\to0$), the term $\hat \s_{AL}^\r$ in Eq.~\eqref{eq:multiNLA} can be discarded, leaving $\ket{\Psi_{\text{N-stage}}}\sim\left(1+\frac \kappa N \hat a\dag\hat b\dag\right)^N\ket{0_{AB}}$. In the limit $N\to\infty$, this state becomes a perfect, non truncated, EPR state:
\begin{equation}
\ket{\Psi_{\text{N-stage}}}\to e^{\kappa  \hat a\dag\hat b\dag}\ket{0_{AB}} \sim \ket{\EPR_{AB}^\k}.
\end{equation}

\begin{figure}[!!h]
\begin{tabular}{rc}
\rotatebox{90}{$\e_{\rm best}$}&
\begin{tabular}{c}
\includegraphics[width=8cm]{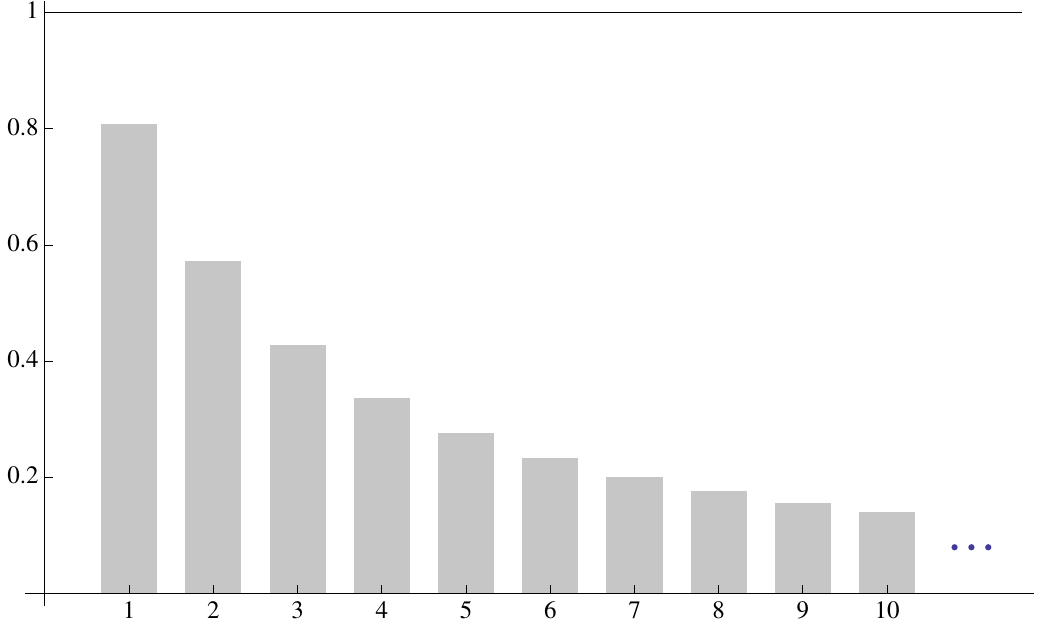}
\end{tabular}  \\
&
$N$
\end{tabular}
\caption[Saturation levels]{In the limit where the probability of success $\Pi$ approaches 0, the best entanglement distillable $\e_{\rm best}$ is obtained by minimising $\e$ over $\k$, for all states of the form $(1+\frac \kappa N \hat a\dag\hat b\dag)^N\ket{0_{AB}}$.  This value depends on the number of stages $N$ in the NLA and tends exponentially slowly towards 0 as $N\to\infty$.}
\label{Saturation}
\end{figure}

In a realistic situation (finite number of stages and non zero success rate), the optimal number of stages one should use depends on the final entanglement strength required. Fig.~\ref{Saturation} shows the best entanglement one can distil as a function of the number $N$ of stages. 

For a fixed probability of success, increasing the number of stages only allows to distil better entanglement in the lower loss regime. For example, the 2-stage NLA would only be useful to distil stronger entanglement for loss between $\sim6$dB and $\sim25$dB (see Fig.~\ref{Comp12}.a). The upper bound is set by the minimum success rate we require (here $\Pi=0.01\%$). Moreover, for a fixed probability of success, the purity of the amplified state will decrease as the number of stages increases. This leads us to conclude that it is beneficial to choose the NLA with a minimal number of stages enabling us to achieve a desired entanglement strength.

Aside from considerations of experimental complexity, another reason to avoid a large number of stages $N$ is that the  maximum entanglement achievable grows exponentially slower with $N$ as shown in Fig.~\ref{Saturation}.

\section{Conclusion}

We have have studied the performances of several different ideal noiseless linear amplifiers based on multiple quantum scissors. We characterized the devices in terms of both the strongest EPR entanglement, and the best entanglement-purity compromise achievable. We have shown that for a fixed amount of loss, a finite entangled resource with a single stage NLA will achieve better performance than an infinitely entangled resource without an NLA.

We found that the maximum entanglement achievable increases with the number of stages used. However, less trivially, we also found that for a fixed entanglement strength and fixed probability of success, the purity of the amplified state decreases as the number of stages increases. Consequently it seems to be beneficial to implement an NLA with a minimal number of stages, in all cases considered here. 

The analysis presented here assumed perfect photon counting and perfect single photon sources. In a real experimental implementation, the detection inefficiency and dark counts in the detectors as well as statistical mixture of vacuum and higher photon numbers in the ancilla modes will inevitably alter the quality of the distillation process by reducing the probability of success but also, more importantly, by degrading the final entanglement due to the presence of other sources of decoherence. While we expect these additional imperfections to diminish the performance of the NLA (in particular, affecting the asymptotic values of entanglement and purity when the loss approaches 0), we expect the inclusion of this imperfection to be linearly incremental and also expect the general trends and conclusions derived here to qualitatively hold. The asymptotic curves will also saturate when the probability of success $\Pi$ tends toward the limit set by the dark count rates.

\section{Aknowledgement}
This research was conducted by the Australian Research Council Centre of Excellence for Quantum Computation and Communication Technology (Project number CE110001027).

%open questions:
%- what happens for a realistic NLA (imperfect detection and photons production)
%- is it better to use many NLA with lower gain or a single NLA with high gain 

%\bibliographystyle{plainbib}
%\bibliography{ref6}

\end{document}